\documentclass[12pt]{article}
\usepackage{amssymb}
\usepackage{amsmath}
\usepackage{amsfonts}
\usepackage{mitpress}

\newdimen\dummy
\dummy=\oddsidemargin
\begin{document}

\begin{center}
\textbf{On lower and upper bounds for Asian-type options: a unified approach}
\begin{equation*}
\end{equation*}%
Alexander Novikov \footnote{%
University of Technology, Sydney. Present address: PO Box 123, Broadway,
Department of Mathematical Sciences, University of Technology, Sydney, NSW
2007, Australia; e-mail:Alex.Novikov@uts.edu.au} and Nino Kordzakhia 
\footnote{%
Macquarie University, Sydney, Australia; e-mail:Nino.Kordzakhia@mq.edu.au}\\[%
0pt]
\end{center}

\begin{equation*}
\end{equation*}%
\textbf{Abstract. } In the context of dealing with financial risk management
problems it is desirable to have accurate bounds for option prices in
situations when pricing formulae do not exist in the closed form. A unified
approach for obtaining upper and lower bounds for Asian-type options,
including options on VWAP, is proposed in this paper. The bounds obtained
are applicable to the continuous and discrete-time frameworks for the case
of time-dependent interest rates. Numerical examples are provided to
illustrate the accuracy of the bounds.\newline
\newline

\textbf{1. Introduction.} We aim to obtain accurate bounds for option prices%
\begin{equation*}
C_{T}=Ee^{-R_{T}}F_{T}(S),
\end{equation*}%
where $R_{t}=\int_{0}^{t}r_{s}ds$, $r_{s}$ is an interest rate, $F_{T}(S)$
is an Asian-type payoff of the option written on the stock price $%
S=(S_{t},0\leq t\leq T),~T$ is the maturity time. (We assume that all random
processes are defined \emph{\ }on the \emph{\ filtered probability space (}$%
\Omega ,\{F_{t}\}_{t\geq 0},P)$).

The typical payoff for Asian-type options is%
\begin{equation}
F_{T}(S)=(\int_{0}^{T}(S_{u}-K)d\mu (u))^{+},  \label{asian}
\end{equation}%
where $x^{+}=\max [x,0]=(-x)^{-}~$for any~$x,~K~$is a fixed strike, $\mu (u)$
is a distribution function on the interval $[0,T].$ Using the notation%
\begin{equation*}
\overline{h}=\int_{0}^{T}h_{u}\mu (du),h\in H,
\end{equation*}%
where $H$ is the class of adapted random processes $h=(h_{s},0\leq s\leq
T)\,\ $such that $\int_{0}^{T}|h_{u}|\mu (du)=\overline{|h|}<\infty ~~a.s.,$
we can rewrite (\ref{asian}) as follows%
\begin{equation}
F_{T}(S)=(\overline{S}-K)^{+}=\overline{(S-K)}^{+}.  \label{asian2}
\end{equation}

In relation to discretely monitored options (\textbf{DMO}) or continuously
monitored options (\textbf{CMO}) the distribution function $\mu $ can be
discrete or continuous respectively. This setup also includes the case of
call options on the {volume-weighted average price (VWAP), that is} 
\begin{equation*}
A_{T}:=\frac{\sum\limits_{t_{j}\leq T}S_{t_{j}}U_{t_{j}}}{%
\sum\limits_{t_{j}\leq T}U_{t_{j}}},~F_{T}(S)=(A_{T}-K)^{+},
\end{equation*}%
where $U_{t_{j}}$ is a traded volume at the moment $t_{j}$.~By setting%
\begin{equation*}
\mu (u):=\frac{\sum\limits_{t_{j}\leq u}U_{t_{j}}}{\sum\limits_{t_{j}\leq
T}U_{t_{j}}}~,0\leq u\leq T,
\end{equation*}%
we obtain the representations (\ref{asian}) and (\ref{asian2}) for options
on VWAP.

Below we develop \ a unified approach to obtaining lower and upper bounds
for Asian-type DMO\ and CMO including VWAP with a general term structure of
interest rate.

The presentation of classical Asian payoffs in the form (\ref{asian}) was
mentioned by Rogers and Shi \cite{RS} and Ve\v{c}e\v{r} (\cite{Vecer}) where
they used the PDE approach for finding $C_{T}$ for CMO under the geometric
Brownian motion (\textbf{gBm}) model and constant interest rates. Thus,
using the notation (\ref{asian}) we can consider an essentially wider class
of options compare to the papers \cite{RS} and (\cite{Vecer}).

The paper \cite{RS} generated a flow of related results about lower and
upper bounds under different settings. We would like to mention here the
pioneering paper by Curran \cite{Cur} and the unpublished paper by Thompson 
\cite{Thom}; in fact, the latter contains some ideas which we are developing
further here. One can find in literature many other similar modifications of
lower and upper bounds, see e.g. (\cite{Albrech}), \cite{Lord} and (\cite%
{March}). We would like to mention the paper by Chen and Lyuu \cite{ChenLyuu}
containing intensive numerical results for CMO under the gBm model, and the
paper by Lemmens et al \cite{Physic} which discusses DMO based on bounds for
geometric Levy processes. In \cite{Physic} comparisons to other approaches
were presented; in particular, among other methods, comparisons to the
recursive integration method developed by Fusai and Meucci \cite{Fusai} and
the method utilising comonotonic bounds (e.g. \cite{V}) were given.

Note that all above cited papers based on the assumption that the interest
rate process is constant. Below we illustrate numerically that for
long-dated contacts the price of the Asian option can be essentially
different if one takes into account a term structure of interest rate.

The case of floating strikes, that is options with the payoff $F_{T}(S)=(%
\overline{S}-S_{T})^{+},$ can be reduced to the case (\ref{asian}) and is
not discussed here.

\textbf{2. \textbf{Lower} and Upper bounds.}

Our main result, which we use for the derivation of lower and upper bounds
below, is given in the following

\textbf{Theorem 1.} \emph{Let }$z$\emph{\ be a real number. Then}%
\begin{equation}
C_{T}=\sup_{z,h\in H}Ee^{-R_{T}}(\overline{S}-K)I\{\overline{h}>z\}
\label{plus0}
\end{equation}%
\begin{equation}
=\inf_{h\in H}Ee^{-R_{T}}\overline{(S-K(1+h-\overline{h}))^{+}}
\label{plus1}
\end{equation}%
\emph{where both supremum and infimum are attained by taking}%
\begin{equation}
h_{u}=S_{u}/K  \label{aa}
\end{equation}%
\emph{and }$z=1.$

\textbf{Proof.~}For any $h\in H~$and $z$ 
\begin{equation*}
(\overline{S}-K)^{+}I\{\overline{h}>z\}=(\overline{S}-K)I\{\overline{h}>z\}+(%
\overline{S}-K)^{-}I\{\overline{h}>z\}
\end{equation*}%
\begin{equation*}
\geq (\overline{S}-K)I\{\overline{h}>z\},
\end{equation*}%
thus we obtain%
\begin{equation}
C_{T}=Ee^{-R_{T}}(\overline{S}-K)^{+}\geq \sup_{z,h\in H}Ee^{-R_{T}}(%
\overline{S}-K)I\{\overline{h}>z\}.  \label{LHD}
\end{equation}%
Since $(\overline{S}-K)^{+}=(\overline{S}-K)I\{\overline{S}/K>1\},~$ the
equalities in (\ref{LHD}) and correspondingly in (\ref{plus0})\emph{\ are
attained }$\ $when $z=1$ and $\overline{h}=\overline{S}/K$.

To prove (\ref{plus1}) we note that for any $h\in H$%
\begin{equation*}
C_{T}=Ee^{-R_{T}}\overline{(S-K)}^{+}
\end{equation*}%
\begin{equation}
=Ee^{-R_{T}}(\overline{S-K(1+h-\overline{h})})^{~+}\leq Ee^{-R_{T}}\overline{%
(S-K(1+h-\overline{h}))^{+}},  \label{RHS1}
\end{equation}%
where the last inequality is due to convexity of $x^{+}.$\emph{\ }This
implies that the $C_{T}$ is not greater than infimum of the RHS of (\ref%
{RHS1})~over $h\in H.~$The equality in (\ref{plus0})\emph{\ is attained}
when $h_{u}=S_{u}/K$ since for the latter case%
\begin{eqnarray*}
\overline{(S-K(1+h-\overline{h}))^{+}} &=&\overline{(S-K(1+S/K-\overline{S}%
/K))^{+}} \\
&=&\overline{(\overline{S}-K)}^{+}=(\overline{S}-K)^{+}.
\end{eqnarray*}

\textbf{Remark 1. }\emph{The proof of this result exploits only the property
of an indicator function for the part (\ref{plus0}), Jensen inequality for
the part (\ref{plus1}) and, of course, these elements were used in many
papers including the cited above. Our main observation consists in noting
that (\ref{plus0})=(\ref{plus1}) and\ both supremum and infimum are attained
on the same function; beside we claim that this is true not only for DMO and
CMO under the gBm model but also for options on stocks with general
structure and this includes the case of VWAP as well.}

Further we use the notation%
\begin{equation*}
X_{t}:=\log (S_{t}/S_{0})
\end{equation*}%
and assume that the discounted process $e^{-R_{t}}S_{t}=S_{0}e^{X_{t}-R_{t}}$
is a martingale with respect to the filtration $\{F_{t}\}_{t\geq 0}$, as
required by the non-arbitrage theory (see e.g. \cite{Kara}).

Theorem 1 implies that for all $h\in H\,\ $the following lower and upper
bounds hold%
\begin{equation}
C_{T}\geq LB0:=S_{0}\sup_{z}Ee^{-R_{T}}(\overline{e^{X}}-\frac{K}{S_{0}})I\{%
\overline{h}>z\},  \label{LB0}
\end{equation}%
\begin{equation}
C_{T}\leq UB0:=S_{0}Ee^{-R_{T}}\overline{(e^{X}-\frac{K}{S_{0}}(1+h-%
\overline{h}))^{+}}.  \label{UB0}
\end{equation}%
To find a process$\ h$ producing accurate bounds we need to take into
account a complexity of calculations of the joint distribution of $(X,h,%
\overline{h}).~$Obviously, the problem can be made computationally
affordable when $h_{u}$ is a linear function of $X_{u}$, that is under the
choice%
\begin{equation*}
h_{u}=a(u)X_{u}+b(u)
\end{equation*}%
with some nonrandom functions $a(u)$ and $b(u)$. Since both inequalities (%
\ref{LB0}) and (\ref{UB0}) are, in fact, equalities when (\ref{aa}) holds,
one may try to match the first moments of $h_{u}$ and $S_{u}/K$~that is to
set 
\begin{equation*}
Eh_{u}=E(S_{u}/K),\ Var(h_{u})=Var(S_{u}/K).
\end{equation*}

In this paper we apply \textbf{another} simple choice with $a(u)=a=const$
and $b(u)=0$ i.e.%
\begin{equation}
h_{u}=aX_{u}  \label{for UB}
\end{equation}%
where the constant $a$ needs to be chosen in the upper bound. For the latter
case we have%
\begin{equation}
C_{T}\geq LB1:=S_{0}\sup_{z}Ee^{-R_{T}}(\overline{e^{X}}-\frac{K}{S_{0}})I\{%
\overline{X}>z\},  \label{LB1}
\end{equation}%
\begin{equation}
C_{T}\leq UB1:=S_{0}\inf_{a}Ee^{-R_{T}}\overline{(e^{X}-\frac{K}{S_{0}}%
(1+aX-a\overline{X}))^{+}}.  \label{UB1}
\end{equation}

Note that the calculation of the lower bound (\ref{LB1}) does not depend on
a choice of the constant $a$.

\textbf{Remark 2. }\emph{The lower bound (\ref{LB1}) was, in fact, used in 
\cite{Thom} for the case of CMO; for the case DMO it was used in \cite%
{ChenLyuu}, both under the gBm model; see other similar bounds e.g. in \cite%
{Lord}. The upper bound (\ref{UB1}) seems to be new.}

\textbf{Remark 3. }\emph{Assuming that }$R=(R_{t},0\leq t\leq T)$\emph{\ and 
}$X=(X_{t},0\leq t\leq T)$\emph{\ are independent processes, we can easily
obtain another lower bound which appears originally in \cite{Cur}:}%
\begin{equation}
C_{T}\geq LB2:=S_{0}Ee^{-R_{T}}\overline{(E(e^{X}|\overline{X})-\frac{K}{%
S_{0}})^{+}}.  \label{LB2}
\end{equation}%
\emph{This bound holds due to the equality }$C_{T}=S_{0}Ee^{-R_{T}}\{E(%
\overline{e^{X}-\frac{K}{S_{0}})}^{~+}|\overline{h})\}$\emph{\ and convexity
of }$x^{+}$\emph{. }

Note that under the additional assumption%
\begin{equation}
g(x):=\overline{E(e^{X}|\overline{X}=x)}~\text{is an increasing function of }%
x,  \label{increase}
\end{equation}%
we have 
\begin{equation*}
LB1\geq LB2.
\end{equation*}%
Indeed, one can see that 
\begin{eqnarray*}
LB2 &=&S_{0}Ee^{-R_{T}}\overline{(E(e^{X}|\overline{X})-\frac{K}{S_{0}})}I\{E%
\overline{(e^{X}|\overline{X})}>\frac{K}{S_{0}}\} \\
&=&S_{0}Ee^{-R_{T}}\overline{(E(e^{X}|\overline{X})-\frac{K}{S_{0}})}I\{%
\overline{X}>g^{-1}(\frac{K}{S_{0}})\},
\end{eqnarray*}%
where $\,g^{-1}\,$is the inverse function. Now$~$it is clear that $LB2$ does
not exceed $LB1$ since one can use the obvious representation%
\begin{equation*}
LB1=S_{0}\sup_{z}Ee^{-R_{T}}\overline{(E(e^{X}|\overline{X})-\frac{K}{S_{0}})%
}I\{\overline{X}>z\}.
\end{equation*}%
It is easy to check that the condition (\ref{increase}) holds in the
classical model where $X$ is a Brownian motion and $r_{t}$ is a nonrandom
function.

\textbf{3. The case of Gaussian returns.}

Here we suppose that the process $X=(X_{u},0\leq u\leq T)\ $is Gaussian. To
simplify the exposition we also suppose that the process $r_{t}$ is
nonrandom. The case of stochastic interest rates which are independent of $%
S_{t},~$can be treated in a similar way.

The pair\emph{~}$(X_{u},\overline{X})$,~obviously, has a Gaussian
distribution with\emph{\ }%
\begin{equation}
Cov(X_{u},\overline{X})=\int_{0}^{T}Cov(X_{u},X_{s})d\mu (s),  \label{13}
\end{equation}%
\begin{equation}
Var(\overline{X})=\int_{0}^{T}\int_{0}^{T}Cov(X_{u},X_{s})d\mu (u)d\mu (s).
\label{14}
\end{equation}%
Below we consider a numerical example which corresponds to the gBm model
with 
\begin{equation*}
X_{u}=R_{u}+\sigma W_{u}-\sigma ^{2}/2\ u,
\end{equation*}%
where $W_{u}$ is a standard Bm.

\textbf{1) Bounds for arithmetic Asian options.}

For the case of DMO we assume that $\mu (u)$ is an uniform discrete
distribution on (0,T] with jumps at points%
\begin{equation*}
u_{i}=\frac{i}{N}T,~i=1,...,N,
\end{equation*}%
where $N$ is the number of time units (e.g. trading days).

From (\ref{13}) we obtain 
\begin{equation*}
\kappa (u_{i}):=cov(W_{u_{i}},\overline{W})=\sum_{j=1}^{N}\min
(u_{i},s_{j})T/N=u_{i}(T-\frac{u_{i}}{2}+\frac{T}{2N}),
\end{equation*}%
\begin{equation*}
V_{N}:=Var(\overline{W})=\frac{T}{3}(1+\frac{3}{2N}+\frac{1}{2N^{2}}).
\end{equation*}%
Note that letting $N\rightarrow \infty \,\ $one can obtain the
characteristics needed for the pricing of CMO as well.

For numerical illustrations and comparisons we consider the set of
parameters $~S_{0}=K=100,~\sigma =0.3,~$the interest rate 
\begin{equation}
r_{s}=0.09(1+c/2\sin (2\pi s)),  \label{rt}
\end{equation}%
where the parameter $c=0~$or$~c=1.$

One can speed up calculations of the bounds using the function~$erfc(x)$.
For example, using the Girsanov transformation we have obtained the
following expression for the lower bound%
\begin{equation*}
LB1=
\end{equation*}%
\begin{equation*}
\frac{e^{-R_{T}}S_{0}}{2T~N}\max_{z}[\sum_{i}e^{R_{u_{i}}}erfc\{\sqrt{V_{N}/2%
}(z-\sigma \kappa (u_{i}))\}-\frac{K}{S_{0}}erfc\{\sqrt{V_{N}/2}z\}].
\end{equation*}%
It takes\ less than a quarter of second with Mathematica for any $\sigma $
to find this lower bound. Computing the upper bounds UB2 is also relatively
fast (up to 7 seconds using Mathematica for fixed $a$) but essentially
slower with use of the command \emph{\ FindMinimum} in Mathematica.\emph{\ }%
The optimal value of $a$ for the upper bound (\ref{UB1}) ~is usually found
in the interval \thinspace $(0.7,1)$. In fact, we found that UB1 with the
choice $a=1$ produces a reasonable accuracy.

In Table 1 the numerical results for LB1 and UB1 obtained with Mathematica
are reported with three decimal digits. We provide the calculated bounds for
two cases $c=0$ and $c=1$ in (\ref{rt}); the results for $c=1$ are formatted
in bold and placed in brackets. As an estimate for the price we consider a
midpoint of the interval $(LB1,UB1)$:%
\begin{equation*}
\hat{C}_{T}=\frac{LB1+UB1}{2}.
\end{equation*}%
The following bound is valid for the relative error of $\hat{C}_{T}:$%
\begin{equation*}
|\hat{C}_{T}/C_{T}-1|100\%=(UB1/LB1-1)50\%.
\end{equation*}%
\begin{equation*}
\end{equation*}%
\newpage \textbf{Table 1.}

===================================

\begin{tabular}{lllll}
$T$ & $N$ & $LB1$ & $UB1$ & $error~\%~for~\hat{C}_{T}$ \\ 
1 & 
\begin{tabular}{l}
10 \\ 
50 \\ 
$\infty $%
\end{tabular}
& 
\begin{tabular}{l}
12.162 \textbf{(12.135)} \\ 
11.782 (\textbf{11.785)} \\ 
11.718 (\textbf{11.741})%
\end{tabular}
& 
\begin{tabular}{l}
12.259 (\textbf{12.239}) \\ 
11.829 (\textbf{11.807}) \\ 
11.731 \textbf{(11.769)}%
\end{tabular}
& 
\begin{tabular}{l}
0.4 (\textbf{0.42}) \\ 
0.1 \textbf{(0.11)} \\ 
0.03 \textbf{(0.11)}%
\end{tabular}
\\ 
&  &  &  &  \\ 
9 & 
\begin{tabular}{l}
10 \\ 
50 \\ 
$\infty $%
\end{tabular}
& 
\begin{tabular}{l}
56.344 \textbf{(60.769)} \\ 
56.073 (\textbf{60.066)} \\ 
56.012 (\textbf{60.014)}%
\end{tabular}
& 
\begin{tabular}{l}
57.233 \textbf{(61.568)} \\ 
56.419 \ (\textbf{60.506)} \\ 
56.146 (\textbf{60.197)}%
\end{tabular}
& 
\begin{tabular}{l}
0.78 \textbf{(0.68)} \\ 
0.3 \textbf{(0.37)} \\ 
0.1\textbf{7} \textbf{(0.15)}%
\end{tabular}%
\end{tabular}

===================================

\begin{equation*}
\end{equation*}

As it might be anticipated, the prices for options with longer maturities
(here $T=9$) depend essentially on a term structure of interest rate.

\textbf{2)\ Bounds on DMO and CMO on VWAP options.}

In \cite{NLK} we applied the method of matching moments for finding
approximations for options on VWAP under the assumption that $S_{t}$ is\ a
gBm and the volume process $\ U_{t}$ is a squared Ornstein-Uhlenbeck process
and assuming that $S_{t}$.and $U_{t}$ are independent,$~r_{t}=r=const.~$The
key point in the approach used in \cite{NLK} was the development of
technique for finding the function%
\begin{equation*}
g=(g_{t}:=E\frac{U_{t}}{\overline{U}},0\leq t\leq T).
\end{equation*}%
Again with the choice of $\ h_{t}=aX_{t},~$ Proposition 1 implies the
following bounds%
\begin{equation*}
C_{T}\geq LB1=S_{0}e^{-rT}\sup_{z}E\overline{~(e^{X}-\frac{K}{S_{0}})g}I\{%
\overline{X}>z\},
\end{equation*}%
\begin{equation*}
C_{T}\leq UB1=S_{0}e^{-rT}\inf_{a}E\overline{~(e^{X}-\frac{K}{S_{0}}(1+aX-a%
\overline{X}))^{+}g},
\end{equation*}%
where the averaging is supposed to be with respect to an uniform discrete or
continuous distribution on $(0,T]$ for DMO or CMO cases respectively.

The method for calculation of the function $g~$suggested in \cite{NLK} is
based on the formula 
\begin{equation*}
g_{t}=\int_{0}^{\infty }{\frac{\partial }{\partial z}}\mathbb{E}\left(
e^{zU_{t}-qV_{T}}\right) \bigg|_{z=0}dq,
\end{equation*}%
which leads to an analytical representation for $g$ for the case under
consideration.

For numerical illustrations we consider the case of CMO~with the following
parameters (to match the related results from Stace (2007), (2007a) who used
a different approach via PDE):%
\begin{equation*}
dS_{t}=0.1~S_{t}dt+\sigma S_{t}dW_{t},~S_{0}=110,T=1,~K=100,
\end{equation*}%
\begin{equation*}
U_{t}=X_{t}^{2},dX_{t}=2(22-X_{t})dt+5dW_{t},X_{0}=22.
\end{equation*}

\textbf{Table 2}

===================================

\begin{tabular}{llll}
$\sigma $ & $LB1$ & $MC~(error)$ & $UB1$ \\ 
0.1 & 14.198 & 14.199 (0.0019) & 14.204 \\ 
0.5 & 19.612 & 19.6406 (0.0083) & 19.650 \\ 
0.8 & 25.591 & 25.642 (0.014) & 25.784%
\end{tabular}

===================================

For Monte Carlo we used 10 million trajectories and 500 discretisation
points.

\begin{equation*}
\end{equation*}

\textbf{Acknowledgment.} The authors thanks to Volf Frishling, Yulia Mishura
and Scott Alexander for useful discussions and to Tim Ling for the help with
calculations for Table 2.%
\begin{equation*}
\end{equation*}

\end{document}